\documentclass[twocolumn,showpacs,preprintnumbers,amsmath,amssymb]{revtex4}
\usepackage{epsfig}
\usepackage{color}
\preprint{LESCO/2013}
\begin{document}
\title{Phase diagram and optical conductivity of La$_{1.8-x}$Eu$_{0.2}$Sr$_x$CuO$_4$}
\author{M. Autore$^1$, P. Di Pietro$^2$,  P. Calvani$^2$,  U. Schade$^3$, S. Pyon$^4$ T. Takayama$^5$, H. Takagi$^{4,5}$, and S. Lupi$^6$, }
\affiliation{$^1$Dipartimento di Fisica and INFN, Universit\`a di Roma La Sapienza, Piazzale Aldo Moro 2, I-00185 Roma, Italy}
\affiliation{$^2$CNR-SPIN and Dipartimento di Fisica, Universit\`a di Roma La Sapienza, Piazzale Aldo Moro 2, I-00185 Roma, Italy}
\affiliation {$^3$Helmholtz Zentrum f\"ur Materialen und Energie, Albert-Einstein-Str.15, D-12489 Berlin, Germany}
\affiliation {$^4$Department of Physics, University of Tokyo, Tokyo 1130033, Japan}
\affiliation {$^5$ Max Planck Institute for Solid State Research, Heisenbergstrasse 1, 70569 Stuttgart, Germany}
\affiliation{$^6$CNR-IOM and Dipartimento di Fisica, Universit\`a di Roma La Sapienza, Piazzale Aldo Moro 2, I-00185 Roma, Italy}
\date{\today}

\begin{abstract}
La$_{1.8-x}$Eu$_{0.2}$Sr$_x$CuO$_4$ (LESCO) is the member of the 214 family which exhibits the largest intervals among the structural, charge ordering (CO), magnetic, and superconducting transition temperatures. By using new dc transport measurements and data in the literature we construct the phase diagram of LESCO between  $x$ = 0.8 and 0.20. This phase diagram has been further probed in ac, by measuring the optical conductivity $\sigma_1(\omega)$ of three single crystals with $x$ = 0.11, 0.125, and 0.16 between 10 and 300 K in order to  associate  the extra-Drude peaks often observed in the 214 family with a given phase. The far-infrared peak we detect in underdoped LESCO is the hardest among them, survives up to room temperature and is associated with charge localization rather than with ordering. At the CO transition for the commensurate doping $x$ = 0.125 instead the extra-Drude peak hardens and a pseudogap opens in $\sigma_1(\omega)$, approximately as wide as the maximum superconducting gap of LSCO. 

\end{abstract}
\maketitle

\section{Introduction}

One of the most interesting issues concerning high-$T_c$ cuprates is the competitive coexistence between superconductivity and charge/magnetic order. This problem dates back to the  discovery,  in 1988 \cite{mooden}, of a drop in $T_c$  at the commensurate doping $x$ = 0.125 in La$_{2-x}$Ba$_{x}$CuO$_4$ (LBCO), and found an explanation with the discovery of commensurate charge and spin order at 1/8 hole doping \cite{tranq_nature}, in form of charge stripes separated by antiferromagnetic, neutral walls.   It became then clear that  static charge and spin ordering  competes in cuprates  with  superconductivity and leads to its partial suppression. Further studies on different members of the 214 family  \cite{fujita_LBCO,tranq_LNSCO,ichikawa} showed  that, upon cooling, such ordering  is preceded at a temperature called $T_{d2}$ (see Table I) by a structural transition from the usual Low-Temperature Orthorhombic (LTO)  to a Low-Temperature Tetragonal (LTT) phase, through a rotation of the oxygen octahedra surrounding the Cu atoms. However, this well established landscape  was deeply modified when inelastic neutron scattering \cite{birgeneau,cheong,yamada} and  infrared spectroscopy  \cite{lucarelli,basov_str,ortolani} found anomalies in the low-temperature response of  the 214 compounds also at incommensurate doping, where no static superlattices were found by conventional diffraction. Such features were attributed to fluctuating spin and charge ordering, and are similar to those produced by static ordering.   In both cases, for example, an extra-Drude peak appears in the  far-infrared (FIR) optical conductivity, which indicates charge localization, and damped spin excitations or "paramagnons" \cite{dean} are detected by inelastic x-ray scattering (RIXS). Fluctuating charge order in cuprates seems then to coexist with, or even to favor, superconductivity, and the intriguing implications of this finding have been widely discussed in the literature (for a review, see, \textit{e. g.}, Ref. \onlinecite{kivelson}). 


\begin{figure}[!b]   
\begin{center}   
\leavevmode    
\epsfxsize=8.6cm \epsfbox {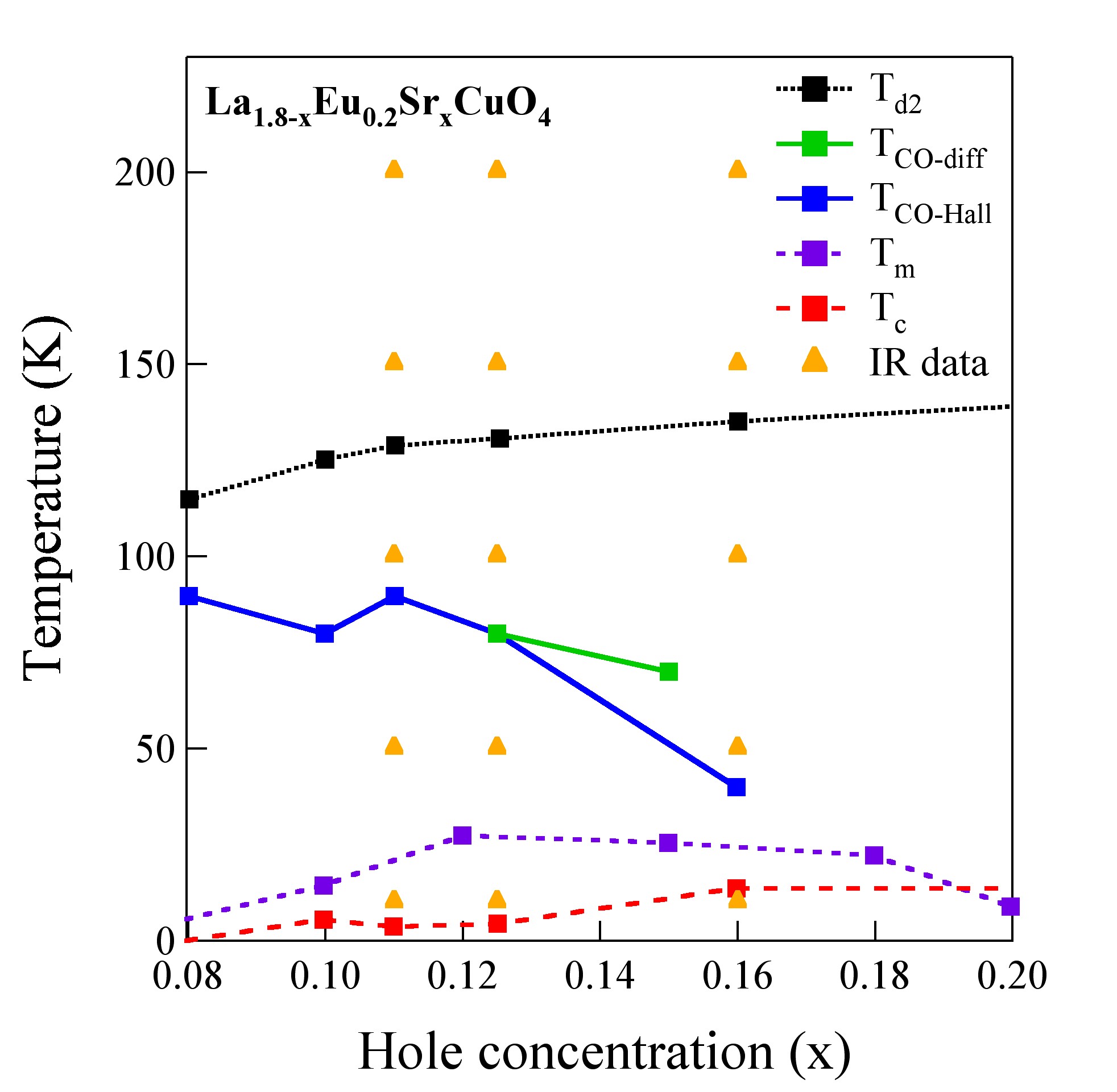}  
\caption{(Color online). Phase diagram of LESCO.  $T_{d2}$ was determined from the anomaly in the temperature derivative of the c-axis resistivity $d\rho_c/dT$, the magnetic ordering temperature $T_m$ was taken from Ref. \onlinecite{klauss}, $T_{CO-diff}$ from Ref. \onlinecite{nature} ($x$ = 0.125) and Ref. \onlinecite {fink} ($x$ = 0.15), and $T_{CO-Hall}$ by the corresponding drop in the Hall coefficient vs. temperature. Yellow triangles indicate the low-$T$ optical measurements presented here.}
\label{phase_diag}
\end{center}
\end{figure}


In an attempt to better understand the interplay between ordering phenomena and superconductivity, another member of the 214 family, La$_{1.8-x}$Eu$_{0.2}$Sr$_x$CuO$_4$ (LESCO) has been investigated in recent years. Therein, below $x$ = 0.17 no Meissner effect can be detected \cite{kataev}, while clear indications of  antiferromagnetic (AF) order were found for $0 \leq x \leq$ 0.014 by muon spin rotation \cite{klauss}. Short-range magnetic order was detected  in the same experiment for $x >$ 0.08. Evidence for stripe formation, with a doping-dependent wavevector,  was provided by Resonant X-ray Diffraction below $T_{CO-diff}$ = 80 K at $x$ = 1/8 and below 65 K at $x$ = 0.15 \cite{fink}. Here also, commensurate order competes with superconductivity, as demonstrated by a recent pump-probe optical experiment. The stripes of LESCO at 1/8 doping were destabilized by a mid-IR pumping of the  Cu-O stretching mode. By then probing in the Terahertz range both the real and imaginary part of the optical conductivity, the authors observed that the superconducting state was restored with a strong increase in $T_c$ \cite{fausti}. 

By using  data in the literature and original dc transport measurements (in particular the c-axis resistivity $\rho_c$ and its derivative $d\rho_c/dT$) we could construct the phase diagram of La$_{1.8-x}$Eu$_{0.2}$Sr$_x$CuO$_4$ between $x$ = 0.08 and 0.20,  as shown in Fig. \ref{phase_diag}. With respect to the leading member of the family, LSCO,  its "superconducting dome" is less pronounced and  coexists with  an antiferromagnetic phase, which survives at higher temperatures. Optimum doping is achieved for $x$ = 0.16,  where static charge ordering  disappears from the X-ray diffraction spectra. As a result of this phase diagram, in comparison with the other charge-ordered  214 compounds, LESCO exhibits the largest separations among the structural, magnetic, and charge ordering transition temperatures (see Table \ref{temps}). This will make easier, in the present optical study, to separate the effects of long-range charge order from those which can be ascribed to the other transitions, or simply to charge localization effects.   Even if  LESCO  has been studied with different techniques for about fifteen years, to date its equilibrium  optical properties have not been reported  to our knowledge.


\begin{table}[h!]
\centering
\begin{tabular}{ccccc}

Compound & $T_{d2}$ (K) & $T_{CO}$ (K) & $T_{M}$ (K) & $T_c$ (K) \\
\\
La$_{1.48}$Nd$_{0.4}$Sr$_{0.12}$CuO$_4$ \cite{tranq_LNSCO} & 70  & 60 & 50 & 2 \\
La$_{1.875}$Ba$_{0.05}$Sr$_{0.075}$CuO$_4$ & 55 & 40 \cite{kim} & 50 & 10 \\
La$_{1.675}$Eu$_{0.2}$Sr$_{0.125}$CuO$_4$ & 125 \cite{klauss}  & 80 \cite{fink} & 45 \cite{hucker} & 5 \\
\\
\end{tabular}
\caption{Transition temperatures of three compounds of the 214 family with 0.125 hole doping.  $T_{d2}$ refers to the structural LTT-LTO transition, $T_{CO}$ and $T_m$ are the temperatures of charge and spin ordering, respectively, and $T_c$ is the superconducting critical temperature. LESCO exhibits the maximum separation among the above temperatures.}
\label{temps}
\end{table}


\section{Experiment and results}
The single crystals of La$_{1.8-x}$Eu$_{0.2}$Sr$_x$CuO$_4$ were all grown at the University of Tokyo using the Travelling-Solvent Floating-Zone (TSFZ) technique and  were fully characterized as reported in Ref. \onlinecite{nature}. The diffraction measurements were performed on the BL19LXU beamline at RIKEN SPring-8, while the resistivity $\rho_c(T)$ and the Hall coefficient $R_H(T)$ of each sample were measured using a standard six-terminal
AC technique \cite{nature}.Three crystals were selected for the optical measurements: two underdoped samples having $x$ = 0.11 (3.0x3.5x0.3 mm in size) and $x$ = 0.125 (2.0x2.5x0.3 mm) and an optimally doped crystal with $x$ =0.16 (8x4x4 mm).  Their transition temperatures are reported in Table \ref{tab_LESCO}. 


\begin{table}[h!]
\centering
\begin{tabular}{ccccc} 
\\
$x$ & $T_{d2}$ (K) & $T_{CO}$ (K) & $T_m$ (K) & $T_{c}$ (K)   \\ 
\\
0.11 & 129 & 90 & 20 &4\\ 
0.125 & 132 & 80 & 30 & 5\\ 
0.16 & 135 & 40 & 25 & 14 \\ 
\\
\end{tabular} 
\caption{Transition temperatures measured in LESCO single crystals with the same compositions as the three samples considered in the present study. For the  employed techniques, refer to the caption of Fig. \ref{phase_diag}.}
\label{tab_LESCO}
\end{table}


The reflectivity $R(\omega)$ of the crystals  was measured at near-normal incidence, in the $ab$ plane, by a Michelson interferometer from 30 to 20000 cm$^{-1}$ in a helium-flow cryostat. The temperature range was 10-300 K, with an error on temperature of $\pm$ 2 K. The reference in the infrared (visible) was a thin gold (silver) layer deposited in situ on the sample. In order to obtain the real part of the optical conductivity $\sigma_1(\omega)$ by standard Kramers-Kronig transformations, $R(\omega)$ was extended to high frequencies using LSCO data from Ref. \onlinecite{tail}, then extrapolated to $\omega = \infty$ by a power law. The extrapolation to zero frequency was instead provided by a Drude-Lorentz fit to the FIR reflectivity. The ($x, T$) positions of the spectra are marked  by triangles in the phase diagram of Fig. \ref{phase_diag}.


\begin {figure}[!b]   
\begin{center}
\leavevmode    
\epsfxsize=8.6cm \epsfbox {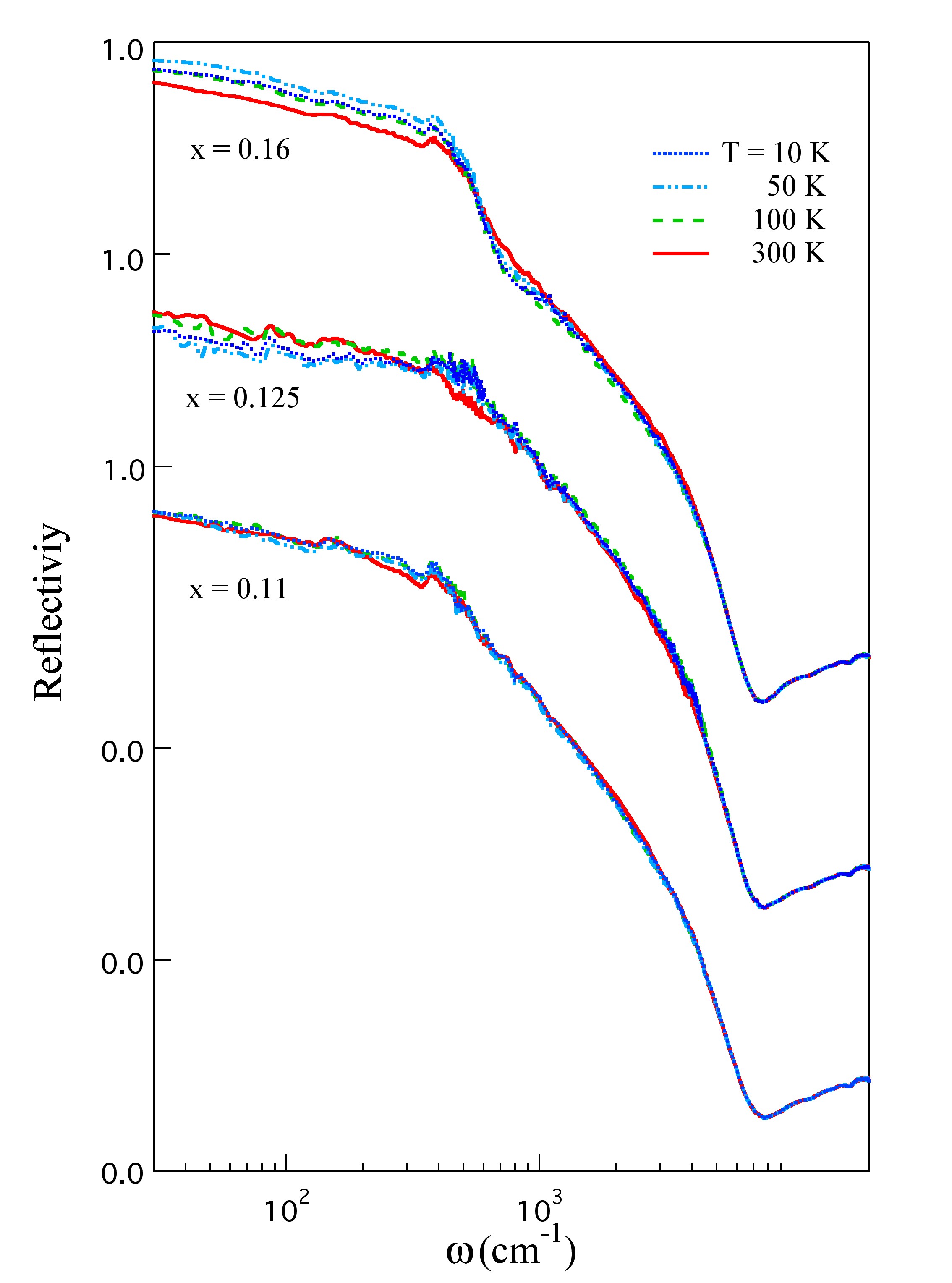}   
 \caption{(Color online). Reflectivity in the $ab$ plane for the three La$_{1.8-x}$Eu$_{0.2}$Sr$_{x}$CuO$_4$ crystals at different temperatures.} 
\label{R}
\end{center}
\end{figure}



\begin {figure}[!b]   
\begin{center}
\leavevmode    
\epsfxsize=8.6cm \epsfbox {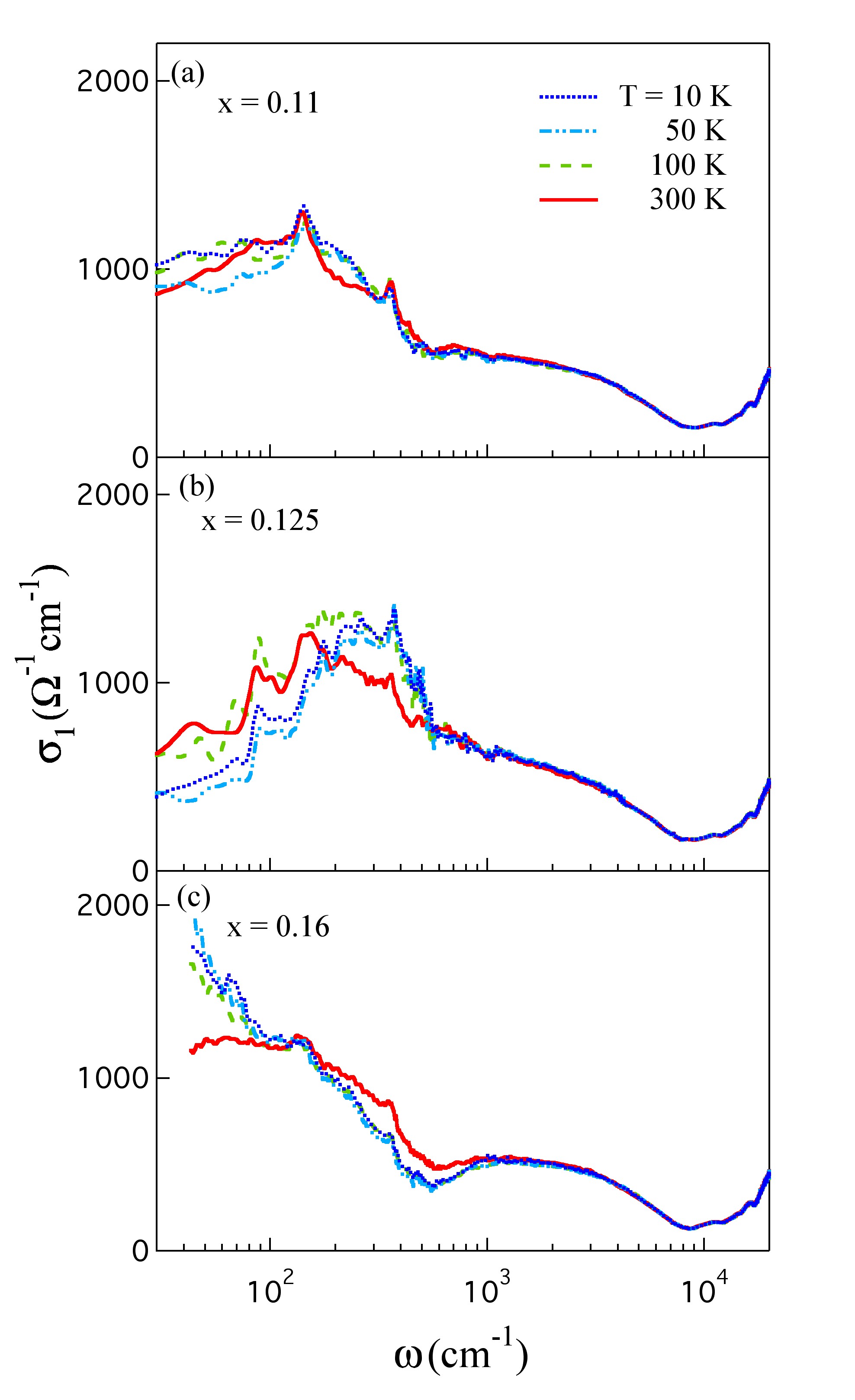}    \caption{(Color online). Optical conductivity in the $ab$ plane for the La$_{1.8-x}$Eu$_{0.2}$Sr$_{x}$CuO$_4$ crystals at different temperatures.} 
\label{sigma}
\end{center}
\end{figure}


The reflectivity spectra are shown in Fig. \ref{R} and their temperature variation is concentrated in the far-infrared range. At higher frequencies, they all exhibit a minimum in the near IR at the so-called \textit{screened plasma frequency} $\tilde{\omega}_p\simeq 7500$ cm$^{-1}$. In metallic cuprates this value cannot be ascribed to the Drude term only, but results from its superposition with  a mid-infrared band (MIR), as discussed below. An interband transition, the Cu-O charge-transfer band, causes the raise in  $R(\omega)$ above $\tilde{\omega}_p$. \\


\begin{figure}[!b]
\begin{center}
\leavevmode    
\epsfxsize=8.6cm \epsfbox {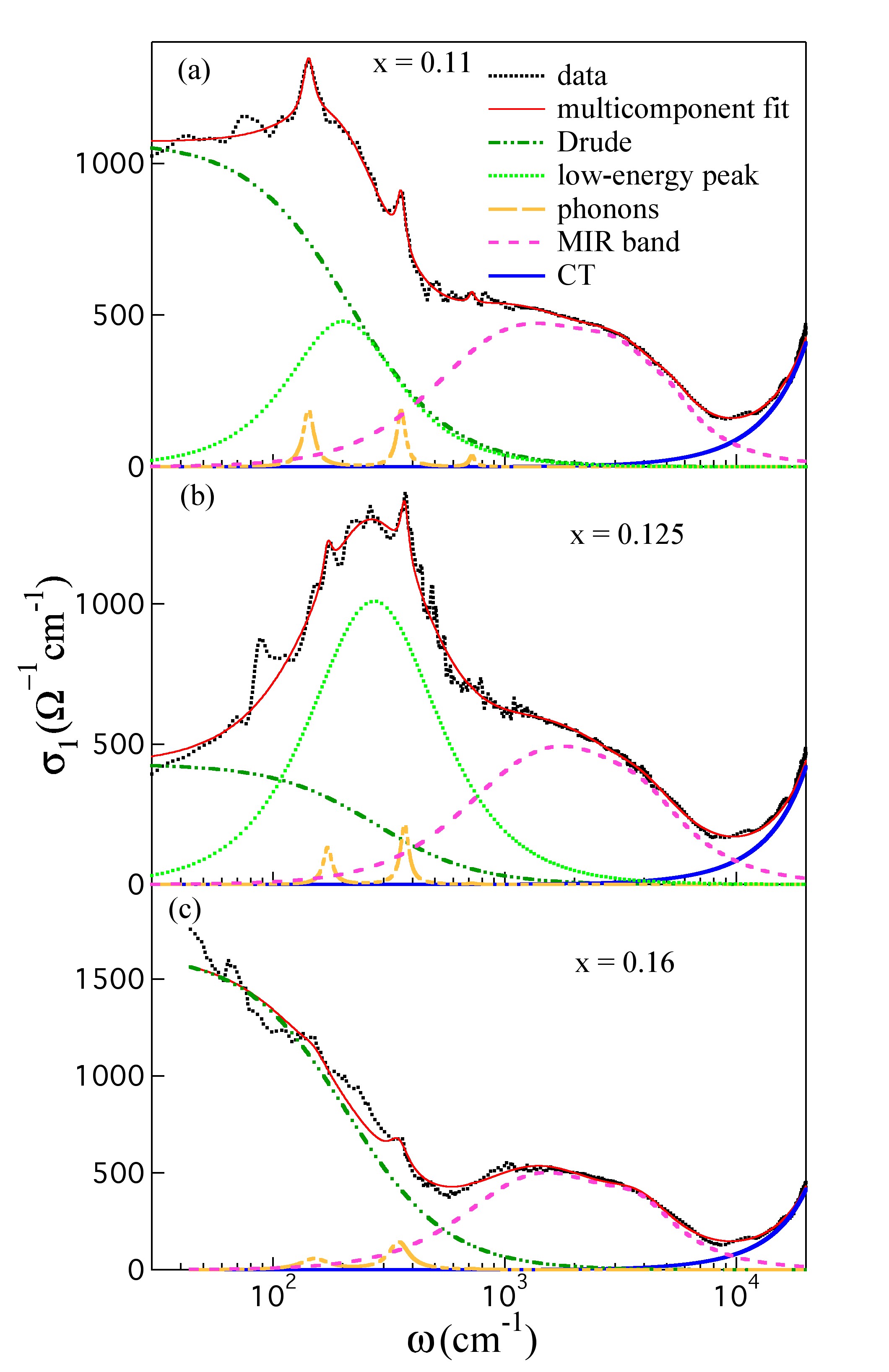}    
\caption{(Color online). Optical conductivity of the La$_{1.8-x}$Eu$_{0.2}$Sr$_{x}$CuO$_4$ crystals  at 10 K (dotted lines), with the results of Drude-Lorentz fits.} 
\label{fit}
\end{center}
\end{figure}


The real part $\sigma_1(\omega)$ of the optical conductivity, extracted from $R(\omega)$ as reported above, is shown for the three samples in Fig. \ref{sigma}. In addition to  the charge-transfer band above $\sim 9000$ cm$^{-1}$, they all exhibit another feature typical of doped cuprates \cite{lupi}, namely, a broad mid-infrared absorption that in LESCO is peaked around $3000$ cm$^{-1}$.This band, which appears in most strongly correlated materials \cite{baldassarre, lupi_NC, Perucchi}, is related to the optical transitions between the lower- and the upper-Hubbard bands to the single particle peak.


\begin{figure}[!b]   
\begin{center}   
\leavevmode    
\epsfxsize=8.6cm \epsfbox {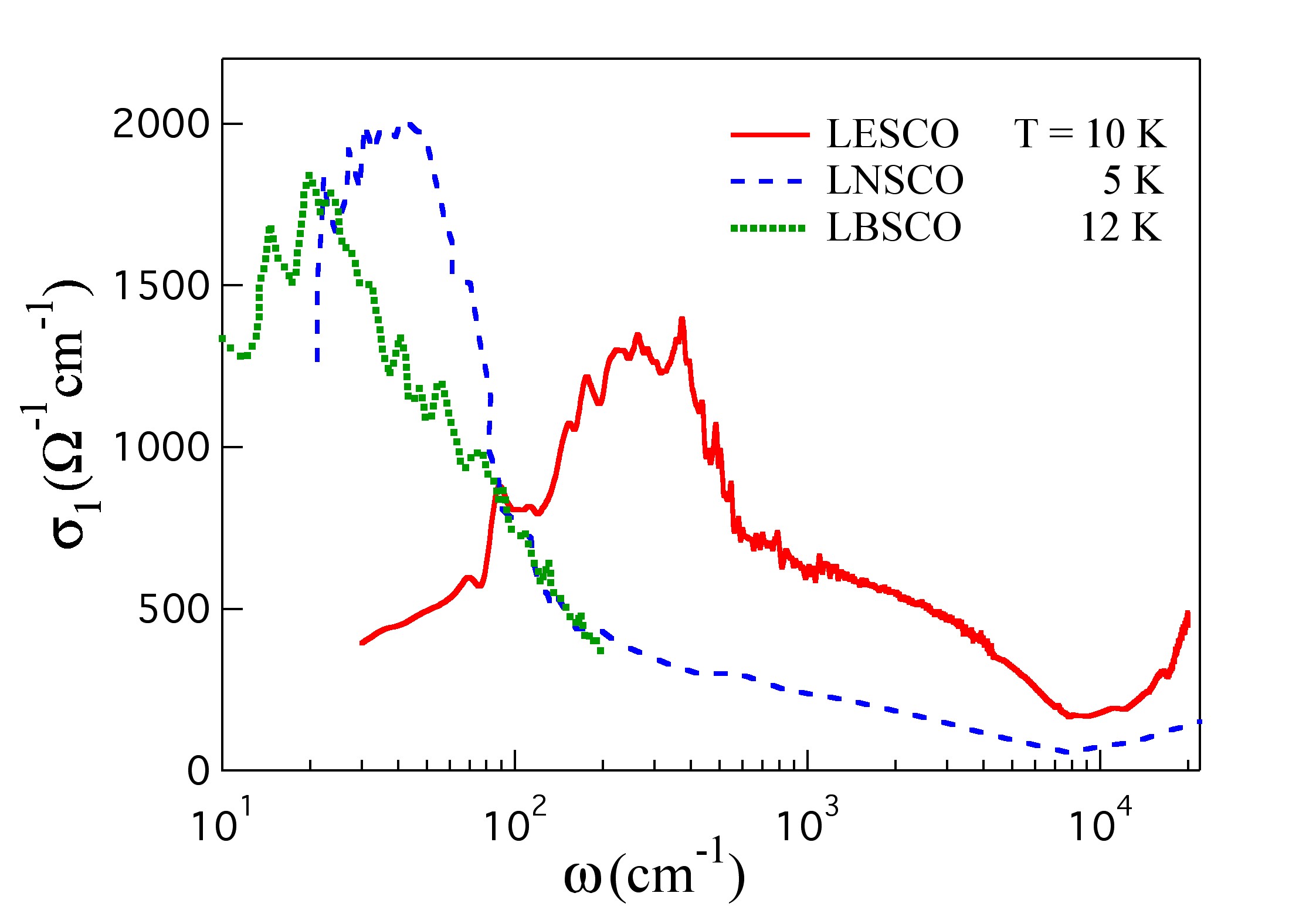}  
\caption{(Color online). Low-temperature optical conductivity, in the CO phase, of  La$_{1.575}$Eu$_{0.2}$Sr$_{0.125}$CuO$_4$, La$_{1.48}$Nd$_{0.4}$Sr$_{0.12}$CuO$_4$ \cite{tranq_LNSCO}, and La$_{1.875}$Ba$_{0.05}$Sr$_{0.075}$CuO$_4$, \cite{kim}. $\sigma_1(\omega)$ in La$_{1.48}$Nd$_{0.4}$Sr$_{0.12}$CuO$_4$ and La$_{1.875}$Ba$_{0.05}$Sr$_{0.075}$CuO$_4$ is divided by a factor 4.} 
\label{comparison}
\end{center}
\end{figure}



\begin{figure}[!b]
\begin{center}
\leavevmode    
\epsfxsize=8.6cm \epsfbox {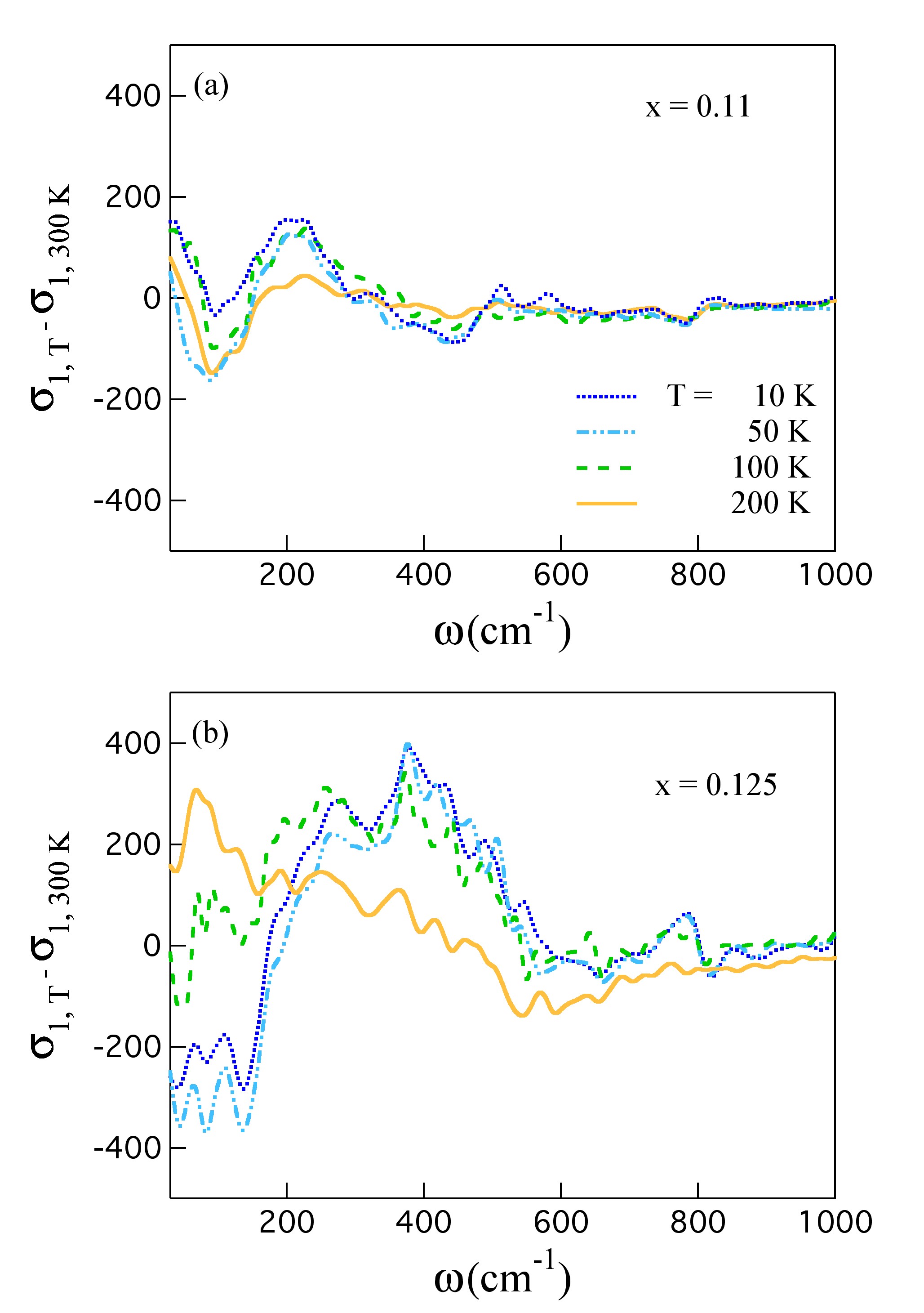}    
\caption{(Color online). Variation of the optical conductivity between different temperatures $T$ and 300 K, for the underdoped  La$_{1.8-x}$Eu$_{0.2}$Sr$_{x}$CuO$_4$ crystals.} 
\label{diff}
\end{center}
\end{figure}


The peculiar features of LESCO appear in the  FIR range, where we will focus our discussion. Therein,  $\sigma_1(\omega)$ exhibits a conventional Drude-like absorption for the  sample at optimum doping ($x$=0.16) in Fig. \ref{sigma}-c. Its Drude peak gradually narrows upon cooling, with a transfer of spectral weight from high to low frequencies around an isosbestic point situated at $\omega \simeq 150$ cm$^{-1}$ and an appreciable decrease above 150 cm$^{-1}$. 
A Drude-Lorentz fit (see below) provides a plasma frequency $\omega_p$ = 4500 cm$^{-1}$, to be compared with $\omega_p$ = 6100 cm$^{-1}$ in Eu-free LSCO. This helps us to understand another observation, namely that $\sigma_1(\omega)$ for $\omega \to 0$ in Fig. \ref{sigma}-c is much lower than for LSCO at optimum doping \cite{lucarelli}. This may be due to a reduction in the charge density, with respect to the Eu-free compound, despite the fact that Eu was considered for a long time isovalent with La. Finally, despite $T_c$ being 14 K, at $T=10 \pm 2$ K the opening of a gap is not observed yet. This is probably due to lack of data below 30 cm$^{-1}$, caused by the small  size of this crystal.

Both in the $x=0.11$ and $x=0.125$ samples, at variance with that at optimum doping,  $\sigma_1(\omega)$ decreases at any $T$ for $\omega \to 0$. The FIR peak responsible for this behavior is clearly shown by the Drude-Lorentz fits  reported at 10 K in Figs. \ref{fit}-a and -b, while it is absent at optimum doping  (Fig. \ref{fit}-c). At $x$ = 0.11 it is  centered around 125 cm$^{-1}$, at $x$ = 0.125  around 170 cm$^{-1}$. The fit also distinguishes the $E_u$ phonon peaks of the 214 $ab$ plane, increasingly shielded for increasing doping, and the broad MIR band reproduced by the sum of two Lorentzians. 
FIR peaks at finite frequencies similar to those in Fig. \ref{fit}-a and -b were observed in LNSCO \cite{basov1}, underdoped LSCO \cite{lucarelli}, and La$_{1.875}$Ba$_{0.125-y }$Sr$_{y}$CuO$_4$ (LBSCO) \cite{ortolani},  and attributed to charge localization and ordering, either static or dynamic.  Figure \ref{comparison} proposes a comparison among the far-infrared conductivities of LESCO (present experiment), LNSCO  \cite{tranq_LNSCO}, and LBSCO \cite{kim}, at commensurate doping and low temperature. All of them show the FIR peak,  but the one in LESCO is found at the highest frequency. Consistently with this finding, in both LESCO samples of  Fig. \ref{sigma} with $x$ =  0.11 and 0.125, the peak survives  at  remarkably high temperatures. They are also higher than all the transition temperatures reported in Table I for $x$ = 0.125 and in the phase diagram of Fig. \ref{phase_diag}. Therefore the charge carriers associated with the FIR peak in these compounds are localized even in the absence of long-range order. For $x = 0.11$, the peak is approximately independent of temperature. For $x = 0.125$ instead, it hardens considerably below the CO transition at 80 K, indicating that it is such shift, not the appearance of the peak itself, which is related to long-range ordering. 

To better follow the process induced by CO in LESCO, we have plotted in Fig. \ref{diff} the variation of the optical conductivity between different temperatures and room temperature for both underdoped  crystals in the far infrared. While at $x=0.11$, as anticipated previously, there is no appreciable change in the whole temperature range, the sample with commensurate doping 
$x=0.125$ exhibits the opening of a pseudogap below 200 cm$^{-1}$. This gap opens at 100 K, just below $T_{d2}=132$ K, and strongly deepens below $T_{CO} = 80$ K, through a transfer of spectral weight towards higher frequencies. One should remark that the direct observation of a pseudogap  in $\sigma_1(\omega)$ is unusual for the $ab$ plane of cuprates, where such phenomena are typically detected in the carrier relaxation rate $\Gamma(\omega)$, after the application of an extended Drude model \cite{timusk}.
Finally, a very interesting result is that the width of the present CO gap (200 cm$^{-1}$) is comparable with that of the optical  gap in superconducting LSCO.

\section{Conclusion}
This work reports an optical investigation of Eu-doped LSCO around the 1/8 commensurate  doping. This member of the 214 family has been selected for its peculiar phase diagram, where the transitions between different structural, electronic, and magnetic phases are well separated in temperature. 
Eu-doping produces in LSCO a pronounced reduction of low-energy conductivity, which basing on the present infrared measurements can be ascribed mainly to a reduction in the plasma frequency and then in the number of the free carriers, despite Eu being considered isovalent with La. In the underdoped samples with $x=0.11$ and $x=0.125$,  a FIR peak is observed at $\omega\sim 200$ cm$^{-1}$ which is similar to those detected in the other 214 compounds which exhibit charge and spin ordering. Nevertheless, here the peak survives at temperatures much higher than both $T_{CO} = 80$ K and $T_{d2} = 132$ K, indicating that it is related to charge localization rather then to long-range ordering. Only in the commensurate $x=0.125$ sample we find evidence for a strong effect below $T_{CO}$. This is the opening of a pseudogap 200 cm$^{-1}$ wide, caused by a pronounced hardening of the above peak, with a recovery of the sum rule around 600 cm$^{-1}$.The pseudogap energy value is impressively close to that for the maximum superconducting gap in the 214 family. This similarity in the energy scale between  commensurate charge/spin ordering and superconductivity substantiates the observation that the former effect can compete with the Cooper pair formation energy and  therefore dramatically reduce $T_c$ in the cuprates of the 214 family.


\begin{thebibliography}{99}
\bibitem{mooden} A. R. Moodenbaugh, Y. Xu,  M. Suenaga, T. J. Folkerts, and R. N. Shelton, Phys. Rev. B \textbf{38}, 4596 (1988).
\bibitem{tranq_nature} J. M. Tranquada,  B. J. Sternlieb, J. D. Axe, Y. Nakamura, and S. Uchida, Nature (London) 375, 561 (1995).
\bibitem{fujita_LBCO} M. Fujita, H. Goka, K. Yamada, J. M. Tranquada, and L. P. Regnault, Phys. Rev. B \textbf{70}, 104517 (2004).
\bibitem{tranq_LNSCO} J. M. Tranquada, J. D. Axe, N. Ichikawa, A. R. Moodenbaugh, Y. Nakamura, and S. Uchida, Phys. Rev. Lett. \textbf{78}, 338 (1997).
\bibitem{ichikawa} N. Ichikawa, S. Uchida, J. M. Tranquada, T. Niem\"oller, P. M. Gehring, S.-H. Lee, and J. R. Schneider, Phys. Rev. Lett. \textbf{85}, 1738 (2000).
\bibitem{birgeneau} R. J. Birgeneau , Y. Endoh, K. Kakurai, Y. Hidaka, T. Murakami, M. A. Kastner, T. R. Thurston, G. Shirane, K. Yamada, Phys. Rev. B \textbf{39}, 2868 (1989). 
\bibitem{cheong} S-W. Cheong, G. Aeppli, T. E. Mason, H. Mook, S. M. Hayden, P. C. Canfield, Z. Fisk, K. N. Clausen, and J. L. Martinez, Phys. Rev. Lett. \textbf{67}, 1791 (1991). 
\bibitem{yamada} K. Yamada, C. H. Lee, K. Kurahashi, J. Wada, S. Wakimoto, S. Ueki, H. Kimura, Y. Endoh, S. Hosoya, G. Shirane, R. J. Birgeneau, M. Greven, M. A. Kastner, and Y. J. Kim, Phys. Rev. B \textbf{57}, 6165 (1998).
\bibitem{lucarelli} A. Lucarelli, S. Lupi, M. Ortolani, P. Calvani, P. Maselli, M. Capizzi, P. Giura, H. Eisaki, N. Kikugawa, T. Fujita, M. Fujita, and K. Yamada Phys. Rev. Lett. \textbf{90}, 037002 (2003).
\bibitem{basov_str} W. J. Padilla, M. Dumm, Seiki Komiya, Yoichi Ando, and D. N. Basov, Phys. Rev. B \textbf{72}, 205101(2008).
\bibitem{ortolani} M. Ortolani, P. Calvani, S. Lupi, U. Schade, A. Perla, M. Fujita, and K. Yamada, Phys. Rev. B \textbf{73}, 184508 (2006).
\bibitem{dean} M. P. M. Dean, G. Dellea, M. Minola, S. B. Wilkins, R. M. Konik, G. D. Gu, M. Le Tacon, N. B. Brookes, F. Yakhou-Harris, K. Kummer, J. P. Hill, L. Braicovich, and G. Ghiringhelli, Phys. Rev. B \textbf{88}, 020403 (2013).
\bibitem{kivelson} S. A. Kivelson, I. P. Bindloss, E. Fradkin, V. Oganesyan, J. M. Tranquada, A. Kapitulnik, and C. Howald
Rev. Mod. Phys. \textbf{75}, 1201 (2003), and Refs. therein.
\bibitem{kataev} V. Kataev, B. Rameev, A. Validov, B. B\"uchner, M. H\"ucker, and R. Borowski, Phys. Rev. B \textbf{58}, R11876 (1998).
\bibitem{klauss} H.-H. Klauss, W. Wagener,  M. Hillberg, W. Kopmann, H. Walf, F. J. Litterst, M. H\"ucker, and B. B\"uchner, Phys. Rev. Lett. \textbf{85}, 4590 (2000).
\bibitem{fink} J\"org Fink, Enrico Schierle, Eugen Weschke, Jochen Geck, David Hawthorn, Viktor Soltwisch, Hiroki Wadati, Hsueh-Hung Wu, Hermann A. D\"urr, Nadja Wizent, Bernd B\"uchner, and George A. Sawatzky, Phys. Rev. B \textbf{79}, 100502 (2009).
\bibitem{fausti} D. Fausti, R. I. Tobey, N. Dean,S. Kaiser, A. Dienst, M. C. Hoffmann, S. Pyon, T. Takayama, H. Takagi, A. Cavalleri, Science \textbf{331}, 189 (2011).
\bibitem{kim} Y.-J. Kim, G. D. Gu, T. Gog, and D. Casa, Phys. Rev. B \textbf{77}, 064520 (2008).
\bibitem{hucker} M. H\"ucker et al., Physica C \textbf{460-462}, 170 (2007). 
\bibitem{nature} O. Cyr-Choinie\'re, R. Daou, F. Laliberte\', D. LeBoeuf, N. Doiron-Leyraud, J. Chang, J.-Q. Yan, J.-G. Cheng, J.-S. Zhou, J. B. Goodenough, S. Pyon, T. Takayama, H. Takagi, Y. Tanaka, and L. Taillefer, Nature \textbf{458}, 743 (2009).
\bibitem{lupi} S. Lupi, P. Maselli, M. Capizzi, P. Calvani, P. Giura, and P. Roy, Phys. Rev. Lett. \textbf{83}, 4852 (1999).
\bibitem{baldassarre} L. Baldassarre, A. Perucchi, D. Nicoletti, A. Toschi, G. Sangiovanni, K. Held, M. Capone, M. Ortolani,
L. Malavasi, M. Marsi, P. Metcalf, P. Postorino, and S. Lupi, Phys. Rev. B \textbf{77}, 113107 (2008).
\bibitem{lupi_NC} S. Lupi, L. Baldassarre, B. Mansart, A. Perucchi, A. Barinov, P. Dudin, E. Papalazarou, F. Rodolakis,
J. -P. Rueff, J. -P. Itie, S. Ravy, D. Nicoletti, P. Postorino, P. Hansmann, N. Parragh, A. Toschi,T. Saha-Dasgupta, O. K. Andersen, G. Sangiovanni, K. Held, and M. Marsi, Nature Communications \textbf{3}, 652 doi:10.1038/ncomms1397 (2010).
\bibitem{Perucchi} A. Perucchi, C. Marini, M. Valentini, P. Postorino, R. Sopracase, P. Dore, P. Hansmann, O. Jepsen, G. Sangiovanni, A. Toschi, K. Held, D. Topwal, D. D. Sarma, and S. Lupi, Phys. Rev. B \textbf{80} , 073101 (2009).
\bibitem{tail} S. Uchida, T. Ido, H. Takagi, T. Arima Y. Tokura, and S. Tajima, Phys. Rev. B \textbf{43}, 7942 (1990).
\bibitem{basov1} M. Dumm, D. N. Basov, Seiki Komiya, Yasushi Abe, and Yoichi Ando, Phys. Rev. Lett. \textbf{88}, 147003 (2002).
\bibitem{timusk}  D. N. Basov and T. Timusk,  Rev. Mod. Phys. \textbf{77} 721 (2005).
\bibitem{cavalleri} D. Fausti, R. I. Tobey, N. Dean, S. Kaiser, A. Dienst, M. C. Hoffmann, S. Pyon, T. Takayama, H. Takagi and A. Cavalleri, Science \textbf{331}, 189 (2011).

\end{thebibliography}
\end{document}